\documentclass{PoS}

\usepackage{amsmath}
\DeclareMathOperator{\pslash}{\displaystyle{\not}}

\title{Hidden Analytic Structure of Higgs Amplitudes and Maximal Transcendentality Principle}

\ShortTitle{Hidden Analytic Structure of Higgs Amplitudes and Maximal Transcendentality Principle}

\author{Qingjun Jin\\
        Graduate School of China Academy of Engineering Physics, \\ No. 10 Xibeiwang East Road, Haidian District, Beijing, 100193, China\\
        E-mail: \email{qjin@gscaep.ac.cn}}

\author{\speaker{Gang Yang}\\
        CAS Key Laboratory of Theoretical Physics, Institute of Theoretical Physics, \\Chinese Academy of Sciences, Beijing 100190, China\\
        School of Physical Sciences, University of Chinese Academy of Sciences,  \\No. 19A Yuquan Road, Beijing 100049, China\\
        E-mail: \email{yangg@itp.ac.cn}}

\abstract{We present the computation of two-loop Higgs plus three-parton amplitudes with dimension-seven operators in Higgs effective field theory. The computation is based on the combination of unitarity cut and integration by parts methods in an unconventional way. The analytic results take remarkably simple form. In particular, the results show that the QCD and ${\cal N}=4$ SYM results share the same leading transcendental parts. This generalizes the so-call maximal transcendentality principle to the Higgs amplitudes with high dimension operators and also with fundamental external quark states. Further simplicity also exists in lower transcendental parts, suggesting hidden structures beyond maximal transcendentality.}

\FullConference{14th International Symposium on Radiative Corrections (RADCOR2019)\\ 
9-13 September 2019\\
		Palais des Papes, Avignon, France}

\begin{document}

\section{Introduction}

Tremendous progress has been made in the study of scattering amplitudes in the last decades. One main driven force is the existence of surprising simplicity of amplitudes. 
The most famous example is the Parke-Taylor formula for maximally helicity violating (MHV) tree amplitudes \cite{Parke:1986gb}.
Another example is that the six-point two-loop MHV amplitude in the planar ${\cal N}=4$ Yang-Mills theory (SYM) can be simplified in a few lines of classical polylogarithms \cite{Goncharov:2010jf}, based on the computation in \cite{DelDuca:2010zg}. 
Such kind of simplicity is totally unexpected from the traditional Feynman diagram point of view. 
They strongly suggest that there should be some alternative way to understand quantum field theory, both conceptually and methodologically.

As a close cousin of quantum chromodynamics (QCD), the maximally supersymmetric $\mathcal{N}=4$ SYM theory has been an important testing ground and at the center of many of these developments. 
For example, the on-shell unitarity method \cite{Bern:1994zx,Britto:2004nc} and BCFW recursion relation \cite{Britto:2004ap}, initiated in the study of ${\cal N}=4$, have now important applications in computing multijet processes at the Large Hadron Collider (LHC), see e.g. \cite{Berger:2010zx, Bern:2013gka}. 
There are also a few direct connections between ${\cal N}=4$ SYM theory and QCD. First, at tree-level, the gluons amplitudes are equivalent in the two theories. Furthermore, through supersymmetric decomposition \cite{Bern:1994cg}, one-loop ${\cal N}=4$ amplitudes are useful building blocks in one-loop QCD amplitudes. 

More intriguingly, a direct connection between ${\cal N}=4$ SYM and QCD also exists at high loop orders. This is known as the maximal transcendentality principle, which is a (conjectured) correspondence that the maximally transcendental parts are equal for certain quantities in the two theories.
This was first observed in the seminal work \cite{Kotikov:2002ab, Kotikov:2004er} that the ${\cal N}=4$ twist-2 anomalous dimensions can be obtained from the QCD anomalous dimensions \cite{Moch:2004pa}. Later the correspondence was found in \cite{Brandhuber:2012vm} between that two-loop form factor in ${\cal N}=4$ and QCD Higgs amplitudes with operator ${\rm tr}(F^2)$ involved \cite{Gehrmann:2011aa}, which generalizes the correspondence from pure numbers to kinematics-dependent functions. 
Other evidence of the correspondence was known for Wilson lines \cite{Li:2014afw, Li:2016ctv}.
Our recent work \cite{Jin:2018fak, Jin:2019ile, Jin:2019opr}, on which the present article is based on, generalized the maximal transcendentality principle to Higgs plus three-parton amplitudes with high dimension operators and with quark external states. Related studies can be found in \cite{Brandhuber:2017bkg, Brandhuber:2018xzk, Brandhuber:2018kqb, Banerjee:2017faz, Ahmed:2019nkj}.

\section{Setup: Higgs effective theory}

The physical quantities we will focus are the QCD corrections of Higgs to three-parton amplitudes. 
They have phenomenological relevance to the LHC and future collider experiments, where for probing potential new physics beyond the Standard Model as well as understanding the details of Higgs physics, the high precision computation of Higgs process is mandatory. 
The dominant Higgs production at the LHC is the gluon fusion through a top quark loop \cite{Ellis:1975ap, Georgi:1977gs}. 
With full top mass dependence, the NNLO QCD correction will require a three-loop computation.
A very useful approximation is that, when the top mass $m_{\rm t}$ is much larger than Higgs mass $m_{\rm H}$, the computation can be greatly simplified using an effective field theory (EFT) where the top quark is integrated out \cite{Wilczek:1977zn, Shifman:1979eb, Dawson:1990zj, Djouadi:1991tka, Kniehl:1995tn}.
The Higgs effective Lagrangian can be given as
\begin{equation}
{\cal L}_{\rm eff} = \hat{C}_0H \mathcal{O}_0 + {1\over m_{\rm t}^2} \sum_{i=1}^4 \hat{C}_i H \mathcal{O}_i + {\cal O}\left( {1\over m_{\rm t}^4} \right) \,,
\label{eq:HiggsEFT}
\end{equation}
where $H$ is the Higgs field, $O_0 = {\rm tr}(F^2)$, and the subleading terms contain dimension-6 operators \cite{Buchmuller:1985jz, Gracey:2002he, Neill:2009tn, Harlander:2013oja, Dawson:2014ora}
\begin{align}
\mathcal{O}_{1} & = {\rm tr}(F_\mu^{~\nu} F_\nu^{~\rho} F_\rho^{~\mu}) \,, \\ 
\mathcal{O}_{2} & = {\rm tr}(D_\rho F_{\mu\nu} D^\rho F^{\mu\nu} ) \,,\\
\mathcal{O}_{3} & = {\rm tr}(D^\rho F_{\rho\mu} D_\sigma F^{\sigma\mu}) \,,\\
\mathcal{O}_{4} & = {\rm tr}(F_{\mu\rho} D^\rho D_\sigma F^{\sigma\mu}) \,.
\end{align}
The two-loop Higgs plus three-parton amplitudes with leading ${\cal O}_0$ operator were computed in \cite{Gehrmann:2011aa}.
When the Higgs transverse momentum is comparable to the top mass, the contribution of higher dimension operators in the Higgs EFT will be important. This has been taken into account so far only at NLO QCD accuracy, including the finite top mass effect \cite{Lindert:2018iug,Jones:2018hbb,Neumann:2018bsx}.
We will obtain the new two-loop QCD corrections for Higgs plus 3-parton amplitudes with dimension-7 operators.

Let us mention a few properties of the operators. First, the operators satisfy the linear relation:
\begin{align}
{\cal O}_{2} = {1\over2}\, \partial^2{\cal O}_{0} -4\, g_{\textrm{\tiny YM}} \, {\cal O}_{1} +2\, {\cal O}_{4} \,,
\label{eq:ope-linear-relation}
\end{align}
which can be proved using Bianchi identity  (see e.g. \cite{Gracey:2002he}). Second,  with the equation of motion $D^\mu F_{\mu\nu} \sim g (\bar\psi \gamma_\nu T^a \psi)$, one has 
\begin{align}
\mathcal{O}_3  \sim {\cal O}_3' =(\bar\psi \gamma^\mu T^a \psi) (\bar\psi \gamma_\mu T^a \psi) \,, \qquad \mathcal{O}_4  \sim {\cal O}_4' =F_{\mu\nu}^a D^\mu (\bar\psi \gamma^\nu T^a \psi) \,.
\end{align}
Note that in the pure YM sector, since $D^\mu F_{\mu\nu}=0$, the operators ${\cal O}_3$ and ${\cal O}_4$ will give zero amplitudes. 
For the later discussion, it is convenient to classify operators according to their color structure and length. We introduce a diagrammatic notation: the blob {\includegraphics[height=.45cm]{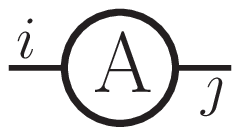}} represents an adjoint field and {\includegraphics[height=.45cm]{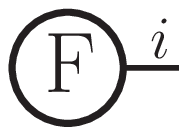}} a fundamental field. By contracting color indices they form a color-singlet operator. The number of blobs is called the length of the operator. For example:
\begin{align}
\text{Length-}2: \quad &
\begin{tabular}{c}{\includegraphics[height=.54cm]{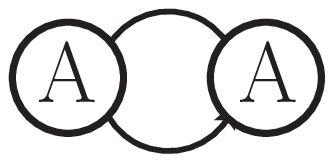}}\end{tabular} \hskip -.2cm:  \ {\cal O}_0 ={\rm tr}(F^2) \,, {\cal O}_2 \, ; \quad 
\begin{tabular}{c}{\includegraphics[height=.5cm]{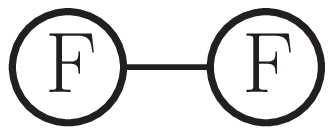}}\end{tabular} \hskip -.2cm: \ \bar\psi\psi \,;  \\
\text{Length-}3: \quad &
\ \begin{tabular}{c}{\includegraphics[height=.85cm]{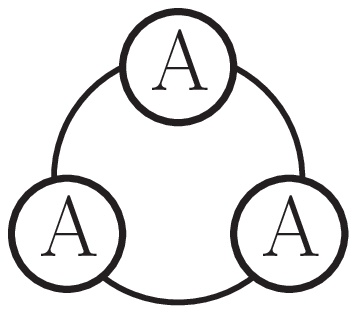}}\end{tabular} \hskip -.2cm:  \ {\cal O}_1 ={\rm tr}(F^3) \, ; \quad \begin{tabular}{c}{\includegraphics[height=.48cm]{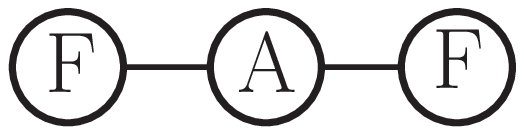}}\end{tabular} \hskip -.2cm: \ F_{\mu\nu}D^\mu(\bar\psi\gamma^\nu\psi) \sim \mathcal{O}_4 \,.
\label{eq:oper-examples}
\end{align}
Note that $\mathcal{O}_3$ is a length-4 operator, and its minimal non-zero tree form factor requires four external quarks $\mathcal{O}_3\rightarrow qq\bar{q}\bar{q}$.

For terminology, we note that the Higgs amplitudes with $n$ partons is equivalent to the form factor with an operator ${\cal O}_i$ in the EFT \eqref{eq:HiggsEFT}:
\begin{equation}
{\cal F}_{{\cal O}_i,n} = \int d^4 x \, e^{-i q\cdot x} \langle p_1, \ldots, p_n | {\cal O}_i(x) |0 \rangle \,, \qquad q^2 = m_H^2 \,.
\end{equation}
Therefore, we will also refer Higgs amplitudes as form factors. 
The operator relation \eqref{eq:ope-linear-relation} implies a relation for the form factors:
\begin{align}
\label{eq:O1-linear-relation}
{\cal F}_{{\cal O}_{2}} = {1\over2}\, q^2 \,{\cal F}_{{\cal O}_{0}} -4\, g_{\textrm{\tiny YM}} \, {\cal F}_{{\cal O}_{1}} +2\, {\cal F}_{{\cal O}_{4}} \,.
\end{align}

\section{Computation: Unitarity-IBP}

Our computation is based on a new strategy of combining the generalized unitarity method \cite{Bern:1994zx, Bern:1994cg, Britto:2004nc} and integration by parts (IBP) reduction \cite{Chetyrkin:1981qh, Tkachov:1981wb} in an unconventional way. 

Unitarity method provides a powerful tool to construct loop integrand from their physical singularities. This is usually used to construct loop amplitudes or form factors at integrand level, where by applying cuts (i.e. setting internal propagators to be on-shell), the loop integrand factorizes into a product of tree results. Using simple tree building blocks, one can reconstruct the loop integrand more efficiently comparing to the traditional Feynman diagram method. 
After the full integrand is obtained, it can be further reduced to a small set of master integrals via IBP. 
This procedure can be summarized as:
\begin{equation}
\mathcal{F}^{(l)}\Bigr|_{\rm cut} \xrightarrow{\ {\rm reconstruction}\ }\mathcal{F}^{(l)}=\sum_a I_a\xrightarrow{\ \text{IBP}\ } \sum_i c_iM_i\ ,
\end{equation}
where $M_i$ are IBP master integrals.

An improved strategy that we use is to apply IBP directly on the cut integrand. Integrals which are not permitted by the cut are set to zero during IBP reduction. In this way, one computes the coefficients of IBP master integrals that contribute in certain cut channel separately. It is important to note that although we apply cuts in the IBP, the coefficients of the cut-permitted master integrals are the final complete coefficients, since the integrals set to zero during cut-IBP never reduce to the cut-permitted integrals. Then by collecting all possible cuts, one obtain the full amplitudes or form factors. This strategy can be summarized as:
\begin{equation}
\mathcal{F}^{(l)}\Bigr|_{\rm cut} 
\xrightarrow{\ \text{cut-IBP}\ } \sum_{\text{cut permitted } M_i} c_iM_i\xrightarrow{\ \text{collect all cuts}\ }\sum_i c_iM_i\ .
\end{equation}

We would like to stress several advantages of the new strategy. 
First, unlike the common unitarity construction, it avoids reconstructing the full integrand, which is usually a non-trivial task in particular when non-planar topologies are involved. Second, it simplifies significantly the IBP reduction, not only because the cut integrand is much simpler than the full integrand, but also that one can apply on-shell condition for cut propagators during IBP. Last but not least, it provides strong self-consistency checks for the computation, since the same master integrals can appear in different cuts and the identification of their coefficients provides cross checks.

A more detailed explanation of the above unitarity-IBP construction can be found in \cite{Jin:2019opr}. See also \cite{Kosower:2011ty, Larsen:2015ped, Ita:2015tya, Georgoudis:2016wff, Abreu:2017xsl, Abreu:2017hqn, Boels:2018nrr, Jin:2018fak, Jin:2019ile}. Similar strategy has been recently used in obtaining three-loop four-gluon amplitudes in pure YM \cite{Jin:2019nya}.

\begin{figure}[tb]
\centering
\includegraphics[scale=0.35]{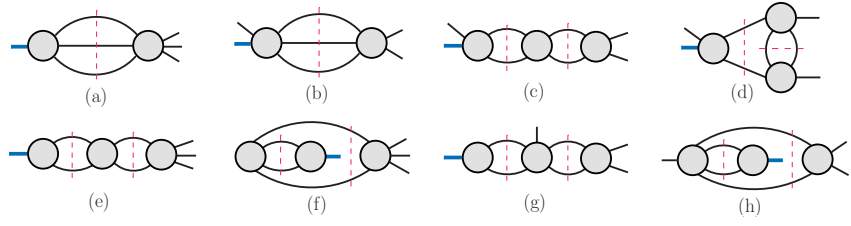}
\caption{The cuts needed in the 2-loop 3-point form factor calculation.}
\label{fig:FF3g2loopAllcuts}
\end{figure}

The planar $D$-dimensional unitarity method is used in the computation of $H\rightarrow 3g$ amplitude. 
The polarization vectors of cut internal gluons can be contracted using the following contraction rule
\begin{equation}
\varepsilon^{\mu}(p) \circ \varepsilon^{\nu}(p)\equiv \sum_{\rm helicities}\varepsilon^{\mu}(p) \varepsilon^{\nu}(p) =\eta^{\mu\nu}
-\frac{q^{\mu}p^{\nu}+q^{\nu}p^{\mu}}{q\cdot p} \,,
\label{eq:helicity-contraction-rule}
\end{equation}
where $q^\mu$ is an arbitrary reference momenta. And similarly for cut integral quark states:
\begin{equation}
u_s(p) \circ \bar{u}_s(p) \equiv \sum_s u_s(p)\bar{u}_s(p)=\pslash{p} \,. 
\label{eq:helicitysum-fermion}
\end{equation}
In the presence of internal quark legs, the two-loop amplitudes contain subleading color contributions. However, these contributions are not intrinsically non-planar, and can still be computed using planar unitarity cuts,  if we assign proper color factors to different internal-state configurations  \cite{Jin:2019opr}. A spanning set of cuts that are enough to determine full two-loop Higgs plus 3-gluon amplitudes are given in Fig.\,\ref{fig:FF3g2loopAllcuts}. 
%
%
\begin{figure}[tb]
\centering
\includegraphics[scale=0.25]{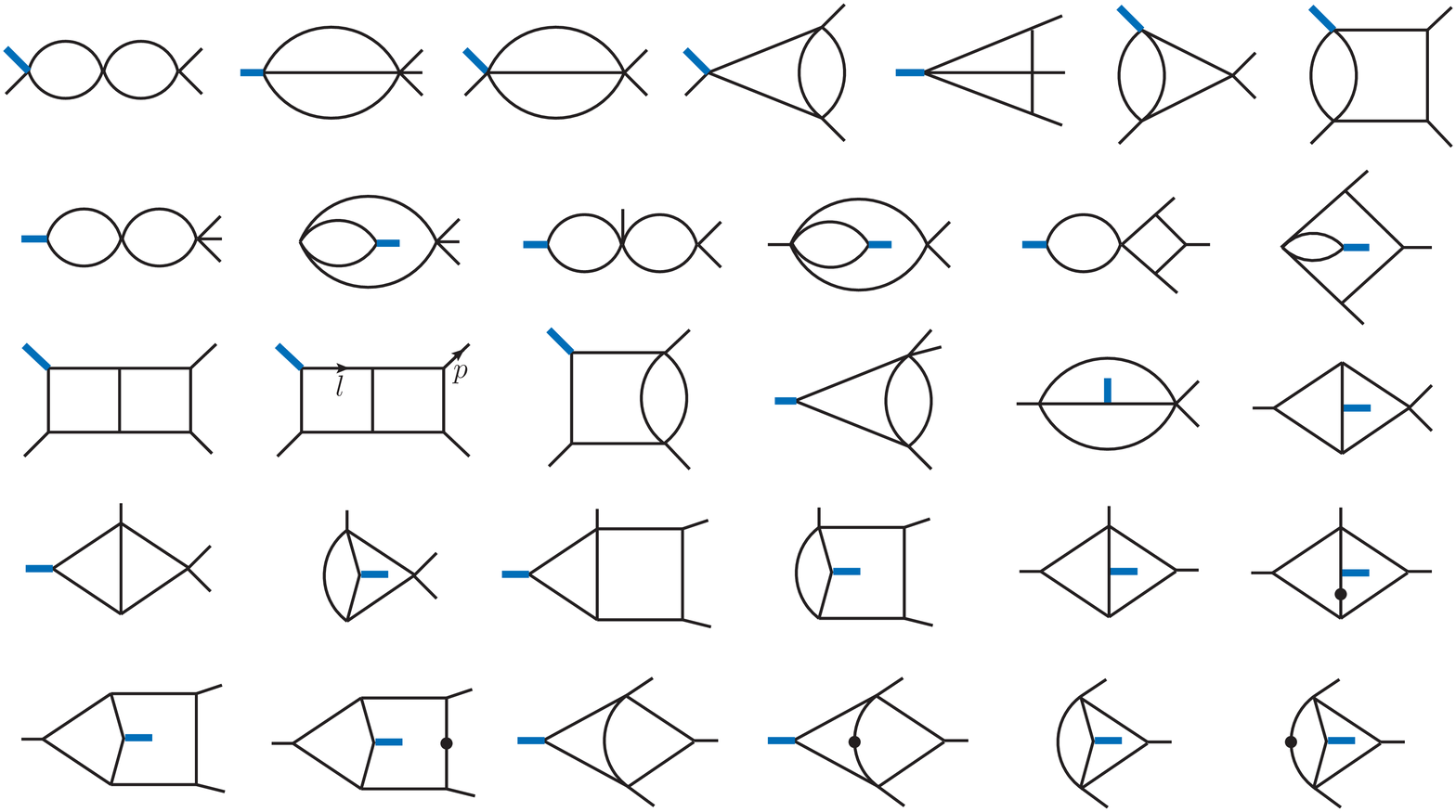}
\caption{The full set of master integrals of the two-loop 3-point form factor. The line with a dot represents a double propagator.}
\label{fig:FF_3g_2loop_MIs}
\end{figure}

The planar unitarity cut alone is not suffice to determine the $H\rightarrow q\bar{q}g$ amplitudes, since they contain contributions which are intrinsically non-planar.
We compute these non-planar contributions with Feynman diagrams  \cite{Hahn:2000kx}. 
It should be possible to compute them by applying non-planar unitarity cuts. 

The tensor reduction of loop integrand is achieved by expanding the integrand in a set of gauge invariant basis $B_{\alpha}$ (see e.g. \cite{Gehrmann:2011aa, Boels:2018nrr}):
\begin{equation}
{\cal F}(\varepsilon_i,p_i,l_a)=\sum_{\alpha} {f}^{\alpha}(p_i, l_a)B_{\alpha}(\varepsilon_i,p_i)  \,,
\label{eq:FprojectedinBandf}
\end{equation}
which can be considered as a gauge invariant implementation of PV reduction. The polarization vectors appear only in the gauge invariant basis, and the loop momenta are contained in the coefficients ${f}^{\alpha}(p_i, l_a)$ which are ready for the IBP reduction.
After IBP, the results are given in terms of master integrals which are list in Fig.\,\ref{fig:FF_3g_2loop_MIs}. These masters have been known in terms of 2d Harmonic polylogarithms \cite{Gehrmann:2000zt,Gehrmann:2001ck}.

\section{Analytic results and hidden structures}

Form factors contain UV and IR divergences, for which we apply dimensional regularization ($D=4-2\epsilon$) in the conventional dimension regularization (CDR) scheme. The UV divergences come from both the coupling constant and the local operator. 
We use the modified minimal subtraction renormalization ($\overline{\rm MS}$) scheme \cite{Bardeen:1978yd}. 
After renormalization, the form factor contains only IR divergences, which take a universal structure \cite{Catani:1998bh}: 
\begin{align}
{\cal F}_{\cal O}^{(1)} &= I^{(1)}(\epsilon) {\cal F}_{\cal O}^{(0)} + {\cal F}_{\cal O}^{(1),{\rm fin}} + {\cal O}(\epsilon) \,,  \\
{\cal F}_{\cal O}^{(2)} &= I^{(2)}(\epsilon) {\cal F}_{\cal O}^{(0)} +  I^{(1)}(\epsilon) {\cal F}_{\cal O}^{(1)} + {\cal F}_{\cal O}^{(2),{\rm fin}} + {\cal O}(\epsilon)   \,,
\end{align}
in which $I^{(l)}$ are known functions independent of operators.

To check the correctness of our results, we have performed several non-trivial checks. First, as we mentioned before, the coefficients of masters are consistent from different unitarity cuts. Second, the $1/\epsilon^m, m=4,3,2$ pole terms at 2-loop are consistent with the universal IR and the 1-loop UV divergences.  Third, our results reproduce known results including the two-loop of Higgs to three-parton amplitudes with the ${\rm tr}(F^2)$ operator \cite{Gehrmann:2011aa}. Finally,  the form factors of different operators satisfy precisely the linear relation \eqref{eq:O1-linear-relation}. 

Since the divergenct terms are well understood, the intrinsic information of the form factor is contained in the finite remainder ${\cal F}_{\cal O}^{(l),{\rm fin}}$. We introduce the normalized remainder $R_{\cal O}^{(l)} = {\cal F}_{\cal O}^{(l),{\rm fin}}/{\cal F}_{\cal O}^{(0),{\rm fin}}$. At two-loop, there are in general six color factors and the remainder can be expanded as:
\begin{align}
R_{{\cal O}}^{(2)} = & N_c^2\, R_{{\cal O}}^{(2),N_c^2} + N_c^0\, R_{{\cal O}}^{(2),N_c^0} + {1\over N_c^2}\, R_{{\cal O}}^{(2),N_c^{-2}}  + {n_f \over N_c}\, R_{{\cal O}}^{(2),n_f/N_c} + N_c\, n_f\, R_{{\cal O}}^{(2),N_c n_f}  + n_f^2\, R_{{\cal O}}^{(2),n_f^2} \,.
\label{eq:fullRem}
\end{align}
We also decompose the remainder according to the transcendentality degree as
\begin{align}
R_{{\cal O}}^{(2)} = \sum_{d=0}^4 R^{(2)}_{{\cal O}; d} ,
\end{align}
where $R^{(2)}_{{\cal O};d}$ has uniform transcendentality degree $d$. At two-loop, the maximal transcendentality degree is 4. Mathematically, transcendental degree characterizes the algebraic complexity of functions and numbers. For instance,  the degree for algebraic numbers or rational functions is zero, $\pi$ or $\log(x)$ has degree $1$, and the Riemann zeta value $\zeta_n$ or polylogrithm ${\rm Li}_n(x)$ has degree $n$. We recall the definition of the polylogrithm:
\begin{equation}
{\rm Li}_n(x) = \sum_{k=1}^\infty {x^k \over k^n} = \int_0^x { {\rm Li}_{n-1}(t) \over t} d t \,, \qquad {\rm Li}_1(x) = - \log(1-x) \,, 
\end{equation}
and $\zeta_n = {\rm Li}_n(1)$.

We consider first the maximal transcendentality (i.e. degree 4) parts. The Higgs amplitudes we consider satisfy the maximal transcendentality principle. The correspondence can be summarized in Table \ref{tab:summary}, which is classified according to the length of operators as well as the type of external particles. 
\begin{table*}[t]
  \begin{center}
  \caption{The universal maximally transcendental properties for Higgs amplitudes or form factors with three partons are summarized. The color-singlet operators are classified according to their lengths and representative examples are provided.}
    \label{tab:summary}
    \begin{tabular}{c|c|c|c|c}
    \hline\hline
     & \multicolumn{2}{c|}{\textbf{Length-2}} &  \multicolumn{2}{c}{\textbf{Length-3}} \\ 
     \hline
     Operators & \begin{tabular}{c}{\includegraphics[height=.5cm]{AA}}\end{tabular} & 
     \begin{tabular}{c}{\includegraphics[height=.5cm]{FF}}\end{tabular} & 
     \begin{tabular}{c}{\includegraphics[height=.9cm]{AAA}}\end{tabular} & 
     \begin{tabular}{c}{\includegraphics[height=.5cm]{FAF}}\end{tabular}
     \\ \hline
     Examples & ${\rm tr}(F^2)$ & $\bar\psi\psi$ &   
     ${\rm tr}(F^3)$ & 
     $F_{\mu\nu}D^\mu(\bar\psi\gamma^\nu\psi)$ 
     \\ \hline
     External Partons & $(g,g,g), (\bar q,q,g)$ & $(\bar q,q,g)$ &     $(g,g,g)$ & $(\bar q, q, g)$   
     \\ \hline
     \begin{tabular}{c} Max. Trans. \\ \textrm{Remainder} \\ (\textrm{with }{$C_F\rightarrow C_A$}) \end{tabular} 
     & \multicolumn{2}{c|}{$R_{\text{len-2};4}(u,v,w)$} 
     & \multicolumn{2}{c}{$R_{\text{len-3};4}(u,v,w)$} 
     \\ \hline\hline
    \end{tabular}
  \end{center}
\end{table*}
The two universal functions $R^{(2)}_{\text{len-2};4}$ and $R^{(2)}_{\text{len-3};4}$ take remarkably simply form:
\begin{align}
R^{(2)}_{\text{len-2};4} = &
-2 \left[ J_4 \left( -\frac{u v}{w}\right)+J_4 \left( -\frac{v w}{u}\right)+J_4 \left( -\frac{w u}{v}\right)\right] 
 -8 \sum_{i=1}^3 \left[ \mathrm{Li}_4 \Big(1-\frac{1}{u_i}\Big)+\frac{\log^4 u_i}{4!} \right] 
 \nonumber\\
& -2 \left[ \sum_{i=1}^3 \mathrm{Li}_2 \Big(1-\frac{1}{u_i}\Big) \right]^2
 +\frac{1}{2} \left[ \sum_{i=1}^3 \log^2 u_i\right]^2 + 2 (J_2^2 - \zeta_2 J_2)  -\frac{\log^4(u v w)}{4!} \nonumber\\
& - \zeta_3 {\log(u v w)} - \frac{123}{8} \zeta_4   ,
\label{eq:RL2}
\end{align}
with
\begin{align}
& J_4(x)  =  \, \mathrm{Li}_4(x)-\log(-x) \mathrm{Li}_3(x)+\frac{\log^2(-x)}{2!} \mathrm{Li}_2(x)  -\frac{\log^3(-x)}{3!} \mathrm{Li}_1(x) - \frac{\log^4(-x)}{48} \,, \\
& J_2  =  \, \sum_{i=1}^3 \bigg( \mathrm{Li}_2(1-u_i)+{1\over2}\log(u_i)\log(u_{i+1}) \bigg) \,,
\end{align}
and
\begin{align}
R^{(2)}_{\text{len-3};4}(u,v,w) := &  -{3\over2} {\rm Li}_4(u) + {3\over4} {\rm Li}_4\left(-{u v \over w} \right) - {3\over4} \log(w) \left[ {\rm Li}_3 \left(-{u\over v} \right) + {\rm Li}_3 \left(-{v\over u} \right)  \right] \nonumber\\
& + {\log^2(u) \over 32} \left[ \log^2(u) + \log^2(v) + \log^2(w) - 4\log(v)\log(w) \right] \nonumber\\
& + {\zeta_2 \over 8} \left[ 5\log^2(u) - 2 \log(v)\log(w) \right]- {1\over4} \zeta_4 + \textrm{perms}(u,v,w) \,,
\label{eq:R2L3-def}
\end{align}
where 
\begin{align}
u = u_1 = {s_{12} \over s_{123} } \,, \ \  v = u_2 = {s_{23} \over s_{123}} \,, \ \  w = u_3 = {s_{13} \over s_{123}} \, .
\end{align}

Let us explain the correspondence in more details. 
First of all, the $n_f$ terms in \eqref{eq:fullRem} do not contain degree-4 parts.
Furthermore, for the Higgs amplitudes with pure external gluons, the $R_{{\cal O}}^{(2),N_c^0}$ and $R_{{\cal O}}^{(2),N_c^{-2}}$ are always zero, due to that full amplitudes can be computed using only planar cuts \cite{Jin:2019opr}. Only $R_{{\cal O}}^{(2),N_c^2}$ contributes to the maximal transcendentality part. They satisfy correspondences:
\begin{align}
R^{(2)}_{{\cal O}_0;4}(1^-, 2^-, 3^\pm) = R^{(2), {\cal N}=4}_{{\cal L}_2;4} = R^{(2)}_{\text{len-2};4}\,, \qquad R^{(2)}_{{\cal O}_1;4}(1^-, 2^-, 3^-) = R^{(2), {\cal N}=4}_{{\cal L}_3;4} = R^{(2)}_{\text{len-3};4}\,,
\end{align}
where ${\cal L}_2$ and ${\cal L}_3$ are the supermultiplet in ${\cal N}=4$ which contain ${\rm tr}(F^2)$ and ${\rm tr}(F^3)$ as a component respectively \cite{Brandhuber:2012vm, Brandhuber:2017bkg, Brandhuber:2018xzk}.
The first relation for ${\cal O}_0 = {\rm tr}(F^2)$ was observed in \cite{Brandhuber:2012vm}. The same $R^{(2)}_{\text{len-2};4}$ function was also obtained in Konishi  form factor in ${\cal N}=4$ SYM \cite{Banerjee:2016kri}. The relation for ${\cal O}_1 = {\rm tr}(F^3)$ was recently shown in \cite{Brandhuber:2017bkg, Jin:2018fak}.

Even more non-trivial correspondence are for the Higgs amplitudes with external quark states, $H\rightarrow q \bar q g$. In such case, the maximal transcendentality terms come from not only the planar but also non-planar components:
\begin{align}
\label{eq:deg4sum}
R_{{\cal O};4}^{(2)}
= N_c^2\, R_{{\cal O}}^{(2),N_c^2} + N_c^0\, R_{{\cal O}}^{(2),N_c^0} + {1\over N_c^2}\, R_{{\cal O}}^{(2),N_c^{-2}} 
= C_A^2\, R_{{\cal O};4}^{(2),C_A^2} + C_A C_F R_{{\cal O};4}^{(2),C_A C_F} + C_F^2 R_{{\cal O};4}^{(2),C_F^2} \,,
\end{align}
where in the second equation we have reorganized the results in terms of the quadratic Casimirs of the adjoint and fundamental representations
\begin{align}
C_A = N_c \,, \qquad C_F = \frac{N_c^2 -1}{2N_c} \,,
\end{align}
respectively.
Each component in \eqref{eq:deg4sum} is a non-trivial function containing non-trivial multiple polylogarithms, however notably, their combination satisfies  \cite{Jin:2019ile}
\begin{align}
& R^{(2), C_A^2}_{{\cal O}_0;4}(1^q, 2^{\bar q}, 3^\pm) + R^{(2),C_A C_F}_{{\cal O}_0;4}(1^q, 2^{\bar q}, 3^\pm) +  R^{(2), C_F^2}_{{\cal O}_0;4}(1^q, 2^{\bar q}, 3^\pm)  = R^{(2)}_{\text{len-2};4}  \,,\\
& R^{(2), C_A^2}_{{\cal O}_4;4}(1^q, 2^{\bar q}, 3^\pm) + R^{(2), C_A C_F}_{{\cal O}_4;4}(1^q, 2^{\bar q}, 3^\pm) + R^{(2), C_F^2}_{{\cal O}_4;4}(1^q, 2^{\bar q}, 3^\pm) = R^{(2)}_{\text{len-3};4}  \,. 
\end{align}
The left hand side of the equations can be understood by setting $C_F \rightarrow C_A$ in the remainder. Similar correspondence was  known for the anomalous dimensions \cite{Kotikov:2002ab, Kotikov:2004er}. As show in  Table \ref{tab:summary}, the same relation also holds if we replace the operator ${\cal O}_0$ to be $\bar\psi\psi$, suggesting this is a property for more general form factors beyond Higgs amplitudes.
For pseudo-scalar Higgs amplitudes with quark states, the universal maximally transcendental part was also noted in \cite{Banerjee:2017faz}.

The Higgs amplitudes contain also non-trivial lower transcendentality parts. Intriguingly, universal structures also exist in the sub-transcendentality parts \cite{Jin:2018fak, Jin:2019opr}. For example, the degree-3 terms of ${\cal O}_1 \rightarrow 3g$ results is
\begin{align}
R^{(2)}_{{\cal O}_1;3} = &   \left( 1+ {u\over w} \right) T_3(u,v,w) + {143\over72}\zeta_3 - {11\over24}\zeta_2\log(u)+ \textrm{perms}(u,v,w) \,,
\end{align}
while the corresponding ${\cal N}=4$ SYM result is given by
\begin{align}
R^{(2), {\cal N}=4}_{{\cal L}_3;3} = & \left( 1+ {u\over w} \right) T_3 + \textrm{perms}(u,v,w) \,,
\end{align}
in which the function $T_3$ is defined as:
\begin{align}
T_3(u,v,w) := & \Big[ -{\rm Li}_3 \left(-{u\over w} \right) + \log(u) {\rm Li}_2\left({v \over 1-u} \right) - {1\over2} \log(1-u) \log\left({w^2\over 1-u}\right) \nonumber\\
& + {1\over2} {\rm Li}_3\left(-{uv \over w}\right) + {1\over2} \log(u)\log(v)\log(w) + {1\over12}\log^3(w) + (u\leftrightarrow u) \Big] \nonumber\\
& +  {\rm Li}_3(1-v) - {\rm Li}_3(u) + {1\over2} \log^2(v) \log\left({1-v\over u}\right) - \zeta_2 \log\left( {u v \over w} \right) \,.
\label{eq:T3-def}
\end{align}
It is interesting to mention that in all QCD results of length-3 operators, the polylogrithm functions, such as ${\rm Li}_3$ and ${\rm Li}_2$, are all combined into $T_3$ functions, what are left are simply $\zeta_3$ or $(\zeta_2\times \log)$ terms. The same $T_3$ function also appeared as building blocks in the ${\cal N}=4$ form factors of more general operators \cite{Loebbert:2015ova, Loebbert:2016xkw, Brandhuber:2017bkg}. 
For transcendentality degree-2 part, one major building block is:
\begin{align}
T_2(u,v) := & \text{Li}_2(1-u)+\text{Li}_2(1-v)+\log (u) \log (v)- \zeta_2 \,.
\label{eq:T2-def}
\end{align}
Explicit expressions for sub-transcendentality parts written in these building blocks can be found in \cite{Jin:2019opr}.

\section{Conclusion}

We compute Higgs amplitudes in Higgs effective theory and study the analytic properties according to their transcendentality structure. For the maximal transcendentality part, we provide further evidences for the maximal transcendentality principle, which applies not only to Higgs amplitudes with high dimension operators but also for the amplitudes with fundamental external quark states. The latter requires to convert the representation of quarks from the fundamental to the adjoint representation:
\begin{align}
& \  \textrm{Max. Tran. of } {(H \rightarrow q \bar q g)}|_{C_F \rightarrow C_A}  \nonumber \\
= & \  \textrm{Max. Tran. of } {(H \rightarrow 3g)} \nonumber \\
= & \  \textrm{Max. Tran. of ${\cal N}=4$ form factors} \nonumber \,.
\end{align}
The explicit correspondences are summarized in Table \ref{tab:summary}, and the universal degree-4 remainders are given in \eqref{eq:RL2} and \eqref{eq:R2L3-def}.
Moreover, simplicities are also observed in the lower transcendentality parts, where the building blocks for degree 3 and 2 parts are given in \eqref{eq:T3-def} and \eqref{eq:T2-def}. 
Similar to the other examples mentioned in the introduction, such simplicities are not obvious at all using standard Feynman diagram and reduction methods. They strongly suggest that there exist alternative ways which can lead to the final simple form more straightforwardly, for which we will report in the future work.

\section*{Acknowledgements}

GY would like to thank the organisers of "RADCOR 2019" for creating a very pleasant and inspiring atmosphere. 
This work is supported in part by the National Natural Science Foundation of China (Grants No. 11822508, 11847612, 11935013),
by the Chinese Academy of Sciences (CAS) Hundred-Talent Program, 
and by the Key Research Program of Frontier Sciences of CAS. 
We also thank the support of the HPC Cluster of ITP-CAS.

\providecommand{\href}[2]{#2}\begingroup\raggedright\endgroup


\begin{thebibliography}{10}

\bibitem{Parke:1986gb}
S.~J. Parke and T.~R. Taylor {\em Phys. Rev. Lett.} {\bf 56} (1986) 2459.

\bibitem{Goncharov:2010jf}
A.~B. Goncharov, M.~Spradlin, C.~Vergu, and A.~Volovich {\em Phys. Rev. Lett.}
  {\bf 105} (2010) 151605, [\href{http://arxiv.org/abs/1006.5703}{{\tt
  arXiv:1006.5703}}].

\bibitem{DelDuca:2010zg}
V.~Del~Duca, C.~Duhr, and V.~A. Smirnov {\em JHEP} {\bf 05} (2010) 084,
  [\href{http://arxiv.org/abs/1003.1702}{{\tt arXiv:1003.1702}}].

\bibitem{Bern:1994zx}
Z.~Bern, L.~J. Dixon, D.~C. Dunbar, and D.~A. Kosower {\em Nucl.Phys.} {\bf
  B425} (1994) 217--260, [\href{http://arxiv.org/abs/hep-ph/9403226}{{\tt
  hep-ph/9403226}}].

\bibitem{Britto:2004nc}
R.~Britto, F.~Cachazo, and B.~Feng {\em Nucl.Phys.} {\bf B725} (2005) 275--305,
  [\href{http://arxiv.org/abs/hep-th/0412103}{{\tt hep-th/0412103}}].

\bibitem{Britto:2004ap}
R.~Britto, F.~Cachazo, and B.~Feng {\em Nucl. Phys.} {\bf B715} (2005)
  499--522, [\href{http://arxiv.org/abs/hep-th/0412308}{{\tt hep-th/0412308}}].

\bibitem{Berger:2010zx}
C.~F. Berger, Z.~Bern, L.~J. Dixon, F.~Febres~Cordero, D.~Forde, T.~Gleisberg,
  H.~Ita, D.~A. Kosower, and D.~Maitre {\em Phys. Rev. Lett.} {\bf 106} (2011)
  092001, [\href{http://arxiv.org/abs/1009.2338}{{\tt arXiv:1009.2338}}].

\bibitem{Bern:2013gka}
Z.~Bern, L.~J. Dixon, F.~Febres~Cordero, S.~Höche, H.~Ita, D.~A. Kosower,
  D.~Maître, and K.~J. Ozeren {\em Phys. Rev.} {\bf D88} (2013), no.~1 014025,
  [\href{http://arxiv.org/abs/1304.1253}{{\tt arXiv:1304.1253}}].

\bibitem{Bern:1994cg}
Z.~Bern, L.~J. Dixon, D.~C. Dunbar, and D.~A. Kosower {\em Nucl. Phys.} {\bf
  B435} (1995) 59--101, [\href{http://arxiv.org/abs/hep-ph/9409265}{{\tt
  hep-ph/9409265}}].

\bibitem{Kotikov:2002ab}
A.~V. Kotikov and L.~N. Lipatov {\em Nucl. Phys.} {\bf B661} (2003) 19--61,
  [\href{http://arxiv.org/abs/hep-ph/0208220}{{\tt hep-ph/0208220}}]. [Erratum:
  Nucl. Phys.B685,405(2004)].

\bibitem{Kotikov:2004er}
A.~Kotikov, L.~Lipatov, A.~Onishchenko, and V.~Velizhanin {\em Phys.Lett.} {\bf
  B595} (2004) 521--529, [\href{http://arxiv.org/abs/hep-th/0404092}{{\tt
  hep-th/0404092}}].

\bibitem{Moch:2004pa}
S.~Moch, J.~A.~M. Vermaseren, and A.~Vogt {\em Nucl. Phys.} {\bf B688} (2004)
  101--134, [\href{http://arxiv.org/abs/hep-ph/0403192}{{\tt hep-ph/0403192}}].

\bibitem{Brandhuber:2012vm}
A.~Brandhuber, G.~Travaglini, and G.~Yang {\em JHEP} {\bf 1205} (2012) 082,
  [\href{http://arxiv.org/abs/1201.4170}{{\tt arXiv:1201.4170}}].

\bibitem{Gehrmann:2011aa}
T.~Gehrmann, M.~Jaquier, E.~Glover, and A.~Koukoutsakis {\em JHEP} {\bf 1202}
  (2012) 056, [\href{http://arxiv.org/abs/1112.3554}{{\tt arXiv:1112.3554}}].

\bibitem{Li:2014afw}
Y.~Li, A.~von Manteuffel, R.~M. Schabinger, and H.~X. Zhu {\em Phys. Rev.} {\bf
  D91} (2015) 036008, [\href{http://arxiv.org/abs/1412.2771}{{\tt
  arXiv:1412.2771}}].

\bibitem{Li:2016ctv}
Y.~Li and H.~X. Zhu {\em Phys. Rev. Lett.} {\bf 118} (2017), no.~2 022004,
  [\href{http://arxiv.org/abs/1604.01404}{{\tt arXiv:1604.01404}}].

\bibitem{Jin:2018fak}
Q.~Jin and G.~Yang {\em Phys. Rev. Lett.} {\bf 121} (2018), no.~10 101603,
  [\href{http://arxiv.org/abs/1804.04653}{{\tt arXiv:1804.04653}}].

\bibitem{Jin:2019ile}
Q.~Jin and G.~Yang \href{http://arxiv.org/abs/1904.07260}{{\tt
  arXiv:1904.07260}}.

\bibitem{Jin:2019opr}
Q.~Jin and G.~Yang \href{http://arxiv.org/abs/1910.09384}{{\tt
  arXiv:1910.09384}}.

\bibitem{Brandhuber:2017bkg}
A.~Brandhuber, M.~Kostacinska, B.~Penante, and G.~Travaglini {\em Phys. Rev.
  Lett.} {\bf 119} (2017), no.~16 161601,
  [\href{http://arxiv.org/abs/1707.09897}{{\tt arXiv:1707.09897}}].

\bibitem{Brandhuber:2018xzk}
A.~Brandhuber, M.~Kostacinska, B.~Penante, and G.~Travaglini {\em JHEP} {\bf
  12} (2018) 076, [\href{http://arxiv.org/abs/1804.05703}{{\tt
  arXiv:1804.05703}}].

\bibitem{Brandhuber:2018kqb}
A.~Brandhuber, M.~Kostacinska, B.~Penante, and G.~Travaglini {\em JHEP} {\bf
  12} (2018) 077, [\href{http://arxiv.org/abs/1804.05828}{{\tt
  arXiv:1804.05828}}].

\bibitem{Banerjee:2017faz}
P.~Banerjee, P.~K. Dhani, and V.~Ravindran {\em JHEP} {\bf 10} (2017) 067,
  [\href{http://arxiv.org/abs/1708.02387}{{\tt arXiv:1708.02387}}].

\bibitem{Ahmed:2019nkj}
T.~Ahmed, P.~Banerjee, A.~Chakraborty, P.~K. Dhani, and V.~Ravindran
  \href{http://arxiv.org/abs/1905.12770}{{\tt arXiv:1905.12770}}.

\bibitem{Ellis:1975ap}
J.~R. Ellis, M.~K. Gaillard, and D.~V. Nanopoulos {\em Nucl. Phys.} {\bf B106}
  (1976) 292.

\bibitem{Georgi:1977gs}
H.~M. Georgi, S.~L. Glashow, M.~E. Machacek, and D.~V. Nanopoulos {\em Phys.
  Rev. Lett.} {\bf 40} (1978) 692.

\bibitem{Wilczek:1977zn}
F.~Wilczek {\em Phys. Rev. Lett.} {\bf 39} (1977) 1304.

\bibitem{Shifman:1979eb}
M.~A. Shifman, A.~I. Vainshtein, M.~B. Voloshin, and V.~I. Zakharov {\em Sov.
  J. Nucl. Phys.} {\bf 30} (1979) 711--716. [Yad. Fiz.30,1368(1979)].

\bibitem{Dawson:1990zj}
S.~Dawson {\em Nucl. Phys.} {\bf B359} (1991) 283--300.

\bibitem{Djouadi:1991tka}
A.~Djouadi, M.~Spira, and P.~M. Zerwas {\em Phys. Lett.} {\bf B264} (1991)
  440--446.

\bibitem{Kniehl:1995tn}
B.~A. Kniehl and M.~Spira {\em Z. Phys.} {\bf C69} (1995) 77--88,
  [\href{http://arxiv.org/abs/hep-ph/9505225}{{\tt hep-ph/9505225}}].

\bibitem{Buchmuller:1985jz}
W.~Buchmuller and D.~Wyler {\em Nucl. Phys.} {\bf B268} (1986) 621--653.

\bibitem{Gracey:2002he}
J.~A. Gracey {\em Nucl. Phys.} {\bf B634} (2002) 192--208,
  [\href{http://arxiv.org/abs/hep-ph/0204266}{{\tt hep-ph/0204266}}]. [Erratum:
  Nucl. Phys.B696,295(2004)].

\bibitem{Neill:2009tn}
D.~Neill \href{http://arxiv.org/abs/0908.1573}{{\tt arXiv:0908.1573}}.

\bibitem{Harlander:2013oja}
R.~V. Harlander and T.~Neumann {\em Phys. Rev.} {\bf D88} (2013) 074015,
  [\href{http://arxiv.org/abs/1308.2225}{{\tt arXiv:1308.2225}}].

\bibitem{Dawson:2014ora}
S.~Dawson, I.~M. Lewis, and M.~Zeng {\em Phys. Rev.} {\bf D90} (2014), no.~9
  093007, [\href{http://arxiv.org/abs/1409.6299}{{\tt arXiv:1409.6299}}].

\bibitem{Lindert:2018iug}
J.~M. Lindert, K.~Kudashkin, K.~Melnikov, and C.~Wever
  \href{http://arxiv.org/abs/1801.08226}{{\tt arXiv:1801.08226}}.

\bibitem{Jones:2018hbb}
S.~P. Jones, M.~Kerner, and G.~Luisoni
  \href{http://arxiv.org/abs/1802.00349}{{\tt arXiv:1802.00349}}.

\bibitem{Neumann:2018bsx}
T.~Neumann \href{http://arxiv.org/abs/1802.02981}{{\tt arXiv:1802.02981}}.

\bibitem{Chetyrkin:1981qh}
K.~Chetyrkin and F.~Tkachov {\em Nucl.Phys.} {\bf B192} (1981) 159--204.

\bibitem{Tkachov:1981wb}
F.~Tkachov {\em Phys.Lett.} {\bf B100} (1981) 65--68.

\bibitem{Kosower:2011ty}
D.~A. Kosower and K.~J. Larsen {\em Phys. Rev.} {\bf D85} (2012) 045017,
  [\href{http://arxiv.org/abs/1108.1180}{{\tt arXiv:1108.1180}}].

\bibitem{Larsen:2015ped}
K.~J. Larsen and Y.~Zhang {\em Phys. Rev.} {\bf D93} (2016), no.~4 041701,
  [\href{http://arxiv.org/abs/1511.01071}{{\tt arXiv:1511.01071}}].

\bibitem{Ita:2015tya}
H.~Ita {\em Phys. Rev.} {\bf D94} (2016), no.~11 116015,
  [\href{http://arxiv.org/abs/1510.05626}{{\tt arXiv:1510.05626}}].

\bibitem{Georgoudis:2016wff}
A.~Georgoudis, K.~J. Larsen, and Y.~Zhang {\em Comput. Phys. Commun.} {\bf 221}
  (2017) 203--215, [\href{http://arxiv.org/abs/1612.04252}{{\tt
  arXiv:1612.04252}}].

\bibitem{Abreu:2017xsl}
S.~Abreu, F.~Febres~Cordero, H.~Ita, M.~Jaquier, B.~Page, and M.~Zeng {\em
  Phys. Rev. Lett.} {\bf 119} (2017), no.~14 142001,
  [\href{http://arxiv.org/abs/1703.05273}{{\tt arXiv:1703.05273}}].

\bibitem{Abreu:2017hqn}
S.~Abreu, F.~Febres~Cordero, H.~Ita, B.~Page, and M.~Zeng
  \href{http://arxiv.org/abs/1712.03946}{{\tt arXiv:1712.03946}}.

\bibitem{Boels:2018nrr}
R.~H. Boels, Q.~Jin, and H.~Luo \href{http://arxiv.org/abs/1802.06761}{{\tt
  arXiv:1802.06761}}.

\bibitem{Jin:2019nya}
Q.~Jin and H.~Luo \href{http://arxiv.org/abs/1910.05889}{{\tt
  arXiv:1910.05889}}.

\bibitem{Hahn:2000kx}
T.~Hahn {\em Comput. Phys. Commun.} {\bf 140} (2001) 418--431,
  [\href{http://arxiv.org/abs/hep-ph/0012260}{{\tt hep-ph/0012260}}].

\bibitem{Gehrmann:2000zt}
T.~Gehrmann and E.~Remiddi {\em Nucl. Phys.} {\bf B601} (2001) 248--286,
  [\href{http://arxiv.org/abs/hep-ph/0008287}{{\tt hep-ph/0008287}}].

\bibitem{Gehrmann:2001ck}
T.~Gehrmann and E.~Remiddi {\em Nucl. Phys.} {\bf B601} (2001) 287--317,
  [\href{http://arxiv.org/abs/hep-ph/0101124}{{\tt hep-ph/0101124}}].

\bibitem{Bardeen:1978yd}
W.~A. Bardeen, A.~J. Buras, D.~W. Duke, and T.~Muta {\em Phys. Rev.} {\bf D18}
  (1978) 3998.

\bibitem{Catani:1998bh}
S.~Catani {\em Phys. Lett.} {\bf B427} (1998) 161--171,
  [\href{http://arxiv.org/abs/hep-ph/9802439}{{\tt hep-ph/9802439}}].

\bibitem{Banerjee:2016kri}
P.~Banerjee, P.~K. Dhani, M.~Mahakhud, V.~Ravindran, and S.~Seth {\em JHEP}
  {\bf 05} (2017) 085, [\href{http://arxiv.org/abs/1612.00885}{{\tt
  arXiv:1612.00885}}].

\bibitem{Loebbert:2015ova}
F.~Loebbert, D.~Nandan, C.~Sieg, M.~Wilhelm, and G.~Yang {\em JHEP} {\bf 10}
  (2015) 012, [\href{http://arxiv.org/abs/1504.06323}{{\tt arXiv:1504.06323}}].

\bibitem{Loebbert:2016xkw}
F.~Loebbert, C.~Sieg, M.~Wilhelm, and G.~Yang {\em JHEP} {\bf 12} (2016) 090,
  [\href{http://arxiv.org/abs/1610.06567}{{\tt arXiv:1610.06567}}].

\end{thebibliography}
\end{document}